\input harvmac
\input epsf

\overfullrule=0pt

\parindent=10pt

%
%
\message{S-Tables Macro v1.0, ACS, TAMU (RANHELP@VENUS.TAMU.EDU)}
%
%
\newhelp\stablestylehelp{You must choose a style between 0 and 3.}%
\newhelp\stablelinehelp{You should not use special hrules when stretching
a table.}%
\newhelp\stablesmultiplehelp{You have tried to place an S-Table inside
another
S-Table.  I would recommend not going on.}%
%
%
\newdimen\stablesthinline
\stablesthinline=0.4pt
\newdimen\stablesthickline
\stablesthickline=1pt
%
%
\newif\ifstablesborderthin
\stablesborderthinfalse
\newif\ifstablesinternalthin
\stablesinternalthintrue
\newif\ifstablesomit
\newif\ifstablemode
\newif\ifstablesright
\stablesrightfalse
%
%
\newdimen\stablesbaselineskip
\newdimen\stableslineskip
\newdimen\stableslineskiplimit
%
%
\newcount\stablesmode
\newcount\stableslines
\newcount\stablestemp
\stablestemp=3
\newcount\stablescount
\stablescount=0
\newcount\stableslinet
\stableslinet=0
%
%
%
\newcount\stablestyle
\stablestyle=0
%
%
\def\stablesleft{\quad\hfil}%
\def\stablesright{\hfil\quad}%
%
%
\catcode`\|=\active%
%
%
\newcount\stablestrutsize
\newbox\stablestrutbox
\setbox\stablestrutbox=\hbox{\vrule height10pt depth5pt width0pt}
\def\stablestrut{\relax\ifmmode%
                         \copy\stablestrutbox%
                       \else%
                         \unhcopy\stablestrutbox%
                       \fi}%
%
%
\newdimen\stablesborderwidth
\newdimen\stablesinternalwidth
\newdimen\stablesdummy
\newcount\stablesdummyc
\newif\ifstablesin
\stablesinfalse
%
%
\def\begintable{\stablestart%
  \stablemodetrue%
  \stablesadj%
  \halign%
  \stablesdef}%
\def\stablesadj{%
  \ifcase\stablestyle%
    \hbox to \hsize\bgroup\hss\vbox\bgroup%
  \or%
    \hbox to \hsize\bgroup\vbox\bgroup%
  \or%
    \hbox to \hsize\bgroup\hss\vbox\bgroup%
  \or%
    \hbox\bgroup\vbox\bgroup%
  \else%
    \errhelp=\stablestylehelp%
    \errmessage{Invalid style selected, using default}%
    \hbox to \hsize\bgroup\hss\vbox\bgroup%
  \fi}%
\def\stablesend{\egroup%
  \ifcase\stablestyle%
    \hss\egroup%
  \or%
    \hss\egroup%
  \or%
    \egroup%
  \or%
    \egroup%
  \else%
    \hss\egroup%
  \fi}%
\def\stablestart{%
  \ifstablesin%
    \errhelp=\stablesmultiplehelp%
    \errmessage{An S-Table cannot be placed within an S-Table!}%
  \fi
  \global\stablesintrue%
  \global\advance\stablescount by 1%
  \message{<S-Tables Generating Table \number\stablescount}%
  \begingroup%
  \stablestrutsize=\ht\stablestrutbox%
  \advance\stablestrutsize by \dp\stablestrutbox%
  \ifstablesborderthin%
    \stablesborderwidth=\stablesthinline%
  \else%
    \stablesborderwidth=\stablesthickline%
  \fi%
  \ifstablesinternalthin%
    \stablesinternalwidth=\stablesthinline%
  \else%
    \stablesinternalwidth=\stablesthickline%
  \fi%
  \tabskip=0pt%
  \stablesbaselineskip=\baselineskip%
  \stableslineskip=\lineskip%
  \stableslineskiplimit=\lineskiplimit%
  \offinterlineskip%
  \def\borderrule{\vrule width \stablesborderwidth}%
  \def\internalrule{\vrule width \stablesinternalwidth}%
  \def\thinline{\noalign{\hrule height \stablesthinline}}%
  \def\thickline{\noalign{\hrule height \stablesthickline}}%
  \def\trule{\omit\leaders\hrule height \stablesthinline\hfill}%
  \def\ttrule{\omit\leaders\hrule height \stablesthickline\hfill}%
  \def\tttrule##1{\omit\leaders\hrule height ##1\hfill}%
  \def\stablesel{&\omit\global\stablesmode=0%
    \global\advance\stableslines by 1\borderrule\hfil\cr}%
  \def\el{\stablesel&}%
  \def\elt{\stablesel\thinline&}%
  \def\eltt{\stablesel\thickline&}%
  \def\elttt##1{\stablesel\noalign{\hrule height ##1}&}%
  \def\elspec{&\omit\hfil\borderrule\cr\omit\borderrule&%
              \ifstablemode%
              \else%
                \errhelp=\stablelinehelp%
                \errmessage{Special ruling will not display properly}%
              \fi}%
  \def\stmultispan##1{\mscount=##1 \loop\ifnum\mscount>3 \stspan\repeat}%
  \def\stspan{\span\omit \advance\mscount by -1}%
  \def\multicolumn##1{\omit\multiply\stablestemp by ##1%
     \stmultispan{\stablestemp}%
     \advance\stablesmode by ##1%
     \advance\stablesmode by -1%
     \stablestemp=3}%
  \def\multirow##1{\stablesdummyc=##1\parindent=0pt\setbox0\hbox\bgroup%
    \aftergroup\emultirow\let\temp=}
  \def\emultirow{\setbox1\vbox to\stablesdummyc\stablestrutsize%
    {\hsize\wd0\vfil\box0\vfil}%
    \ht1=\ht\stablestrutbox%
    \dp1=\dp\stablestrutbox%
    \box1}%
  \def\stpar##1{\vtop\bgroup\hsize ##1%
     \baselineskip=\stablesbaselineskip%
     \lineskip=\stableslineskip%

\lineskiplimit=\stableslineskiplimit\bgroup\aftergroup\estpar\let\temp=}%
  \def\estpar{\vskip 6pt\egroup}%
  \def\stparrow##1##2{\stablesdummy=##2%
     \setbox0=\vtop to ##1\stablestrutsize\bgroup%
     \hsize\stablesdummy%
     \baselineskip=\stablesbaselineskip%
     \lineskip=\stableslineskip%
     \lineskiplimit=\stableslineskiplimit%
     \bgroup\vfil\aftergroup\estparrow%
     \let\temp=}%
  \def\estparrow{\vfil\egroup%
     \ht0=\ht\stablestrutbox%
     \dp0=\dp\stablestrutbox%
     \wd0=\stablesdummy%
     \box0}%
  \def|{\global\advance\stablesmode by 1&&&}%
  \def\|{\global\advance\stablesmode by 1&\omit\vrule width 0pt%
         \hfil&&}%
  \def\vt{\global\advance\stablesmode by 1&\omit\vrule width
\stablesthinline%
          \hfil&&}%
  \def\vtt{\global\advance\stablesmode by 1&\omit\vrule width
\stablesthickline%
          \hfil&&}%
  \def\vttt##1{\global\advance\stablesmode by 1&\omit\vrule width ##1%
          \hfil&&}%
  \def\vtr{\global\advance\stablesmode by 1&\omit\hfil\vrule width%
           \stablesthinline&&}%
  \def\vttr{\global\advance\stablesmode by 1&\omit\hfil\vrule width%
            \stablesthickline&&}%
  \def\vtttr##1{\global\advance\stablesmode by 1&\omit\hfil\vrule width
##1&&}%
  \stableslines=0%
  \stablesomitfalse}
\def\stablesdef{\bgroup\stablestrut\borderrule##\tabskip=0pt plus 1fil%
  &\stablesleft##\stablesright%
  &##\ifstablesright\hfill\fi\internalrule\ifstablesright\else\hfill\fi%
  \tabskip 0pt&&##\hfil\tabskip=0pt plus 1fil%
  &\stablesleft##\stablesright%
  &##\ifstablesright\hfill\fi\internalrule\ifstablesright\else\hfill\fi%
  \tabskip=0pt\cr%
  \ifstablesborderthin%
    \thinline%
  \else%
    \thickline%
  \fi&%
}%
\def\endtable{\advance\stableslines by 1\advance\stablesmode by 1%
   \message{- Rows: \number\stableslines, Columns:  \number\stablesmode>}%
   \stablesel%
   \ifstablesborderthin%
     \thinline%
   \else%
     \thickline%
   \fi%
   \egroup\stablesend%
\endgroup%
\global\stablesinfalse}
%
%

\newcount\no
\figno=0
\def\fig#1#2#3{
\par\begingroup\parindent=0pt\leftskip=1cm\rightskip=1cm\parindent=0pt
\baselineskip=11pt
\global\advance\figno by 1
\midinsert
\epsfxsize=#3
\centerline{\epsfbox{#2}}
\vskip 12pt
{\bf Fig.\ \the\figno: } #1\par
\endinsert\endgroup\par
}
\def\figlabel#1{\xdef#1{\the\figno}}


\lref\GreenSP{
  M.~B.~Green, J.~H.~Schwarz and E.~Witten,
 ``Superstring Theory. Vol. 1: Introduction,''
}

\lref\PolchinskiRQ{
  J.~Polchinski,
  ``String theory. Vol. 1: An introduction to the bosonic string,''
}

\lref\DHokerYR{
  E.~D'Hoker and D.~H.~Phong,
  ``The Box graph in superstring theory,''
  Nucl.\ Phys.\ B {\bf 440}, 24 (1995)
  [arXiv:hep-th/9410152].
}

\lref\SenDP{
A.~Sen,
``How does a fundamental string stretch its horizon?,''
arXiv:hep-th/0411255.
}

\lref\MarcusVS{
  N.~Marcus,
  ``Unitarity And Regularized Divergences In String Amplitudes,''
  Phys.\ Lett.\ B {\bf 219}, 265 (1989).
}

\lref\VerlindeKW{
  E.~P.~Verlinde and H.~L.~Verlinde,
  ``Chiral Bosonization, Determinants And The String Partition Function,''
  Nucl.\ Phys.\ B {\bf 288}, 357 (1987).
}

\lref\WilkinsonTB{
  R.~B.~Wilkinson, N.~Turok and D.~Mitchell,
  ``The Decay Of Highly Excited Closed Strings,''
  Nucl.\ Phys.\ B {\bf 332}, 131 (1990).
}

\lref\MitchellQE{
  D.~Mitchell, N.~Turok, R.~Wilkinson and P.~Jetzer,
  ``The Decay Of Highly Excited Open Strings,''
  Nucl.\ Phys.\ B {\bf 315} (1989) 1
  [Erratum-ibid.\ B {\bf 322} (1989) 628].
}

\lref\DaiCP{
  J.~Dai and J.~Polchinski,
  ``The Decay Of Macroscopic Fundamental Strings,''
  Phys.\ Lett.\ B {\bf 220}, 387 (1989).
}

\lref\IengoGM{
  R.~Iengo and J.~G.~Russo,
  ``Handbook on string decay,''
  JHEP {\bf 0602} (2006) 041
  [arXiv:hep-th/0601072].
}

\lref\ChialvaXM{
  D.~Chialva, R.~Iengo and J.~G.~Russo,
  ``Search for the most stable massive state in superstring theory,''
  JHEP {\bf 0501} (2005) 001
  [arXiv:hep-th/0410152].
}

\lref\ChialvaHG{
  D.~Chialva, R.~Iengo and J.~G.~Russo,
  ``Decay of long-lived massive closed superstring states: Exact results,''
  JHEP {\bf 0312} (2003) 014
  [arXiv:hep-th/0310283].
}

\lref\IengoCT{
  R.~Iengo and J.~G.~Russo,
  ``Semiclassical decay of strings with maximum angular momentum,''
  JHEP {\bf 0303} (2003) 030
  [arXiv:hep-th/0301109].
}

\lref\IengoTF{
  R.~Iengo and J.~G.~Russo,
  ``The decay of massive closed superstrings with maximum angular momentum,''
  JHEP {\bf 0211} (2002) 045
  [arXiv:hep-th/0210245].
}

\lref\OkadaSD{
  H.~Okada and A.~Tsuchiya,
  ``The Decay Rate Of The Massive Modes In Type I Superstring,''
  Phys.\ Lett.\ B {\bf 232}, 91 (1989).
}

\lref\GreenZK{
  M.~B.~Green and G.~Veneziano,
  ``Average Properties Of Dual Resonances,''
  Phys.\ Lett.\ B {\bf 36} (1971) 477.
}

\lref\AmatiFV{
  D.~Amati and J.~G.~Russo,
  ``Fundamental strings as black bodies,''
  Phys.\ Lett.\ B {\bf 454}, 207 (1999)
  [arXiv:hep-th/9901092].
}

\lref\ManesCS{
  J.~L.~Manes,
  ``Emission spectrum of fundamental strings: An algebraic approach,''
  Nucl.\ Phys.\ B {\bf 621}, 37 (2002)
  [arXiv:hep-th/0109196].
}

\lref\SundborgAI{
  B.~Sundborg,
``Selfenergies Of Massive Strings,''
  Nucl.\ Phys.\ B {\bf 319}, 415 (1989).
}

\lref\DHokerMR{
  E.~D'Hoker and D.~H.~Phong,
  ``Dispersion relations in string theory,''
  Theor.\ Math.\ Phys.\  {\bf 98}, 306 (1994)
  [Teor.\ Mat.\ Fiz.\  {\bf 98}, 442 (1994)]
  [arXiv:hep-th/9404128].
}

\lref\susstwo{
L.~Susskind,
  ``Some speculations about black hole entropy in string theory,''
  arXiv:hep-th/9309145.
  }

\lref\polone{G.~T.~Horowitz and J.~Polchinski,
  ``A correspondence principle for black holes and strings,''
  Phys.\ Rev.\ D {\bf 55}, 6189 (1997)
  [arXiv:hep-th/9612146].
}

\lref\poltwo{
 G.~T.~Horowitz and J.~Polchinski,
  ``Self Gravitating Fundamental Strings,''
  Phys.\ Rev.\ D {\bf 57}, 2557 (1998)
  [arXiv:hep-th/9707170].
  }

\lref\venezia{ T.~Damour and G.~Veneziano,
  ``Self-gravitating fundamental strings and black holes,''
  Nucl.\ Phys.\ B {\bf 568}, 93 (2000)
  [arXiv:hep-th/9907030].
  }

\lref\dhabone{A.~Dabholkar, G.~W.~Gibbons, J.~A.~Harvey and F.~Ruiz Ruiz,
  ``SUPERSTRINGS AND SOLITONS,''
  Nucl.\ Phys.\ B {\bf 340}, 33 (1990).
  }

\lref\sentwo{
  A.~Sen,
  ``Extremal black holes and elementary string states,''
  Mod.\ Phys.\ Lett.\ A {\bf 10}, 2081 (1995)
  [arXiv:hep-th/9504147].
  }
\lref\khurione{
  R.~R.~Khuri and R.~C.~Myers,
  ``Dynamics of extreme black holes and massive string states,''
  Phys.\ Rev.\ D {\bf 52}, 6988 (1995)
  [arXiv:hep-th/9508045].
  }

\lref\callanone{
  C.~G.~.~Callan, J.~M.~Maldacena and A.~W.~Peet,
  ``Extremal Black Holes As Fundamental Strings,''
  Nucl.\ Phys.\ B {\bf 475}, 645 (1996)
  [arXiv:hep-th/9510134].
  }

  \lref\mandalone{
  G.~Mandal and S.~R.~Wadia,
  ``Black Hole Geometry Around An Elementary Bps String State,''
  Phys.\ Lett.\ B {\bf 372}, 34 (1996)
  [arXiv:hep-th/9511218].
}

\lref\CornalbaHC{
  L.~Cornalba, M.~S.~Costa, J.~Penedones and P.~Vieira,
  ``From Fundamental Strings to Small Black Holes,''
  arXiv:hep-th/0607083.
}

\lref\DabholkarYR{
  A.~Dabholkar,
  ``Exact counting of black hole microstates,''
  Phys.\ Rev.\ Lett.\  {\bf 94}, 241301 (2005)
  [arXiv:hep-th/0409148].
}

\lref\DabholkarDT{
  A.~Dabholkar, F.~Denef, G.~W.~Moore and B.~Pioline,
  ``Precision counting of small black holes,''
  JHEP {\bf 0510}, 096 (2005)
  [arXiv:hep-th/0507014].
}

\lref\ChialvaKI{
  D.~Chialva and R.~Iengo,
  ``Long lived large type II strings: Decay within compactification,''
  JHEP {\bf 0407}, 054 (2004)
  [arXiv:hep-th/0406271].
}

\lref\MitchellUC{
  D.~Mitchell, B.~Sundborg and N.~Turok,
  ``Decays Of Massive Open Strings,''
  Nucl.\ Phys.\ B {\bf 335}, 621 (1990).
}

\baselineskip 20pt

\Title {\vbox{ \baselineskip12pt
\hbox{hep-th/0607220}\hbox{UCLA/06/TEP/20}}}
{\vbox{ \centerline{Decays  of near BPS heterotic strings}}}

\centerline{ Michael Gutperle and Darya Krym}

\smallskip
\centerline{ Department of Physics and Astronomy, UCLA, Los Angeles,
CA 90095-1547, USA}

                                                                                \vskip .3in \centerline{\bf Abstract}
The decay of highly excited massive string states in compactified heterotic string theories is discussed.
We calculate the decay rate and spectrum of states carrying momentum and winding in the compactified direction.  The longest lived states in the spectrum are near BPS states whose decay is dominated by a single decay channel of  massless radiation which brings the state closer to being  BPS.
\smallskip
\Date{}

\newsec{Introduction}

In general, highly excited strings are unstable and will decay. The
decay process poses interesting questions. What is the lifetime of the unstable string?
What is the decay spectrum? Is the decay dominated by massless
radiation or the splitting of the string into two massive
strings ? These questions have been investigated for various
string theories in the past.

The decay of highly excited open string states was investigated in
\MitchellQE\DaiCP\OkadaSD\MitchellUC. The lifetime of a string state of mass $m_0$
was determined to be $T\sim {1\over  g_s^2} {1\over  m_0}$. This is consistent with the facts, that
an open string has a
constant splitting probability per unit length and the size of an
excited string grows linearly with the mass.

The decay of closed strings is expected to be qualitatively and quantitatively different.
Although a generic closed string state is
expected to behave similarly to the open string \WilkinsonTB\IengoGM ,
the fact, that a closed string can only split (in the absence of
D-branes) when two distinct points of the closed string are coincident, leads
to interesting new phenomena. In a series of papers
\ChialvaXM\ChialvaHG\IengoCT\IengoTF\ChialvaKI,  Iengo and Russo found particular  excited string states, whose lifetime grows with mass like $T={1\over g_s^2}m_0^5$. Therefore, for
large masses, these states are very long lived. Furthermore,
the decay into two massive strings is exponentially suppressed.
The authors explained their results, using a  semiclassical argument,
relating the long lived quantum string state to a classical solution, representing a rotating ring. The string
never self-intersects in this configuration, so that, the dominant decay mode is
through the emission of massless radiation.

Supersymmetry provides a completely different mechanism for the stability of
massive string states.  A BPS state is annihilated by some supersymmetry
charges and  the supersymmetry algebra implies a relation between the mass and the charge of the state. Such BPS states are generically stable against decay (they
can decay into other BPS states at special points of marginal
stability). We ask, whether a state
being near BPS, in the appropriate sense, would also be long lived.

The specific string theory, we are considering, is the
heterotic string compactified on a circle of radius $R$. Due to the difficulty of constructing the vertex operators for
general massive string states, we use a very specialized state, involving
only level 1/2 and level 1 oscillators (i.e. states on the leading Regge trajectories). The set of states is uniquely
parameterized
by the mass and the left and right charges. We compute the decay rate
and the spectrum for such states.

In particular, the compactification allows us to study massive states which are
close to the BPS bound, since  a state
with many left oscillators, but few right oscillators can be constructed.
We show, that these states are, indeed, made long lived by supersymmetry and
that, the dominant decay mode is via the emission of massless radiation.
In addition to the BPS states, satisfying $m_0^2=k_L^2$,
the heterotic string also has extremal states, which satisfy $m_0^2=k_R^2$. We contrast the decays of 'near extremal'
to the 'near BPS'
states.

 Let us briefly review several alternative methods that can be used to analyze the
decays of massive string states before moving on to the approach used in this paper.

First, the three point function $\langle  i\mid V(0) \mid f\rangle$ can be
used to directly calculate the amplitude for a specific decay, where
the string state $i$ changes to $f$, by the emission of a second string
state, represented by the vertex operator $V$.
This method is
particularly efficient for calculating decays involving massless
radiation and extracting the spectrum of the massless radiation.

A variation of this method considers averaging over initial states \GreenZK. The
calculation of   \AmatiFV \ChialvaXM\ManesCS\ starts with the three
point function and considers massless emission only, but sums over all
final states with a fixed mass and averages over the initial states
with a fixed mass. If the masses of the initial string state and the massive decay product are large,
the resulting decay amplitude can be evaluated using a saddle point approximation.  The resulting massless
emission spectrum is that, of a black body, with temperature and
greybody factors depending on the particular string which is
considered.

A second method \DHokerMR\DHokerYR\ extracts the mass shift for a massive string state,
from the residue of the double pole in the $s$-channel of the one loop  scattering amplitude of
four massless states. The analytic continuation in external momenta is well defined for the
four point function and the double pole can be unambiguously identified.

In this paper, we opt to use the optical theorem,
which relates the imaginary part of the
one loop two point function, i.e.  mass shift $\delta m_0^2$ of a string state of mass $m_0$,
to the total
decay rate $\Gamma$ of this state \MarcusVS\SundborgAI\DHokerMR.
\eqn\gamdef{
\Gamma  = {Im(\delta m_0^2)\over 2 m_0}}
The mass shift $\delta m_0^2$ is given by the one loop string amplitude with two  vertex operators associated with the initial massive string state inserted.
\eqn\twoptama{\delta m_0^2=\eqalign{A_2&=   g_s^2  \int d^2\tau \Big\langle V(0) \int d^2z
V^\dagger(z,\bar z)\Big\rangle }}
where $g_s$ is the string coupling constant.
It is possible to extract partial rates for
the decay into two string states of a given mass.
However, the rates are
always inclusive, in the sense that one automatically sums over all
decay products and polarizations of a given mass. The final computation is performed by
expanding the amplitude in powers of
 $e^{2 \pi i \tau}$, $e^{iz}$,$e^{-2 \pi i \bar\tau}$, $e^{-i\bar z}$;
 picking out the contributing monomials using mass and charge conservation;
 analytically evaluating the integrals; and numerically evaluating monomial coefficients.

The relation of black holes and excited fundamental strings goes back to \susstwo\polone\poltwo\venezia.
The exact identification of the (small) black hole entropy and BPS string states
\sentwo\dhabone\khurione\mandalone\callanone\ was achieved using $\alpha'$ corrections to the supergravity action
\SenDP\DabholkarYR\DabholkarDT. BPS black holes have zero temperature and are stable, whereas non BPS black holes do
decay via Hawking radiation. While it is interesting to pursue the relation of the decay of perturbative string states
 and Hawking radiation in light of the string/black hole correspondence, we will limit ourselves to the perturbative
 string in this paper. For some recent work on the relation of absorption in perturbative strings and black holes, see
  \CornalbaHC.

The organization of this paper is as follows. In section 2,
 we discuss the spectrum of the heterotic string, the condition for states to be BPS,
 or near BPS, and construct the vertex operators of the massive string states
 that we will use in the rest of the paper. In section 3,
 we calculate the two point function for these states and extract the decay rate.
 In section 4, we present the results for the decay of near BPS states.
 In section 5, we address the decays of other string states, which are not close to BPS.
 We close with a brief discussion of our results in section 6.

\newsec{BPS and near BPS states in heterotic string theory}

We consider the compactification of the heterotic $SO(32)$  string on a circle of radius $R$. The $S^1$ direction is denoted by  $X^9$ and the   non compact spacetime directions are given by $X^\mu$ with $\mu=0,1,\cdots,8$.
The leftmoving modes make up the ten dimensional superstring and the rightmoving modes make up the 26 dimensional bosonic strings. The sixteen extra chiral bosons are compactified on a $SO(32)$ lattice.
However, we choose not to turn on any Wilson lines along the $S^1$ and to have no oscillators
along either the $S^1$ or heterotic lattice directions.

Since the right and left momenta, on the circle, break up into momentum proper, $l$,
and winding, $m$, the $U(1)$ charges are of the following form (in the absence of Wilson lines)\foot{In these formulas and in the rest  of the paper we have set $\alpha'=2$.},
\eqn\uonecha{k_L={l\over R} + {m R\over 2},  \quad k_R = {l\over R} - {m R\over 2}  }

The physical state condition for the left and the right movers
\eqn\physstate{\eqalign{L_0 \mid state \rangle&=({1\over2}(k^\mu k_\mu+({l\over R}+{mR\over 2})^2)+N)\mid state\rangle={1\over2}\mid state\rangle \cr
\bar{L_0}\mid state\rangle&=({1\over2}(k^\mu k_\mu+({l\over R}-{mR\over 2})^2)+\bar N)\mid state\rangle=1\mid state\rangle}}
 lead to the following expressions for the masses
\eqn\massshell{m_0^2 = k_L^2 + 2N -{1} , \quad m_0^2 = k_R^2 + 2({\bar N}-1) }
From which the level matching constraint follows
\eqn\levelmatch{l m + N-\bar N +{1\over 2}=0}
The supersymmetry algebra of the heterotic string with one compactified direction is
\eqn\susyal{\{ Q_a , Q^\dagger_b\} = 2(C \Gamma_\mu {\cal P}_+)_{ab} k^\mu+ 2(C\Gamma^9{\cal P}_+)_{ab} k_{L}}
Where $k_{L}$ is given by \uonecha.  The supersymmetry algebra \susyal\ implies that  massive states
can  preserve eight of the sixteen supersymmetries if they saturate the BPS bound
\eqn\BPSbound{m_0^2 = k_L^2}
It follows from \massshell\ that a perturbative BPS state has $N={1\over 2}$. For notational convenience it is useful to introduce a shifted leftmoving excitation number $K$ defined by
\eqn\kdea{K=N-{1\over2}}
A BPS state hence has $K=0$ and non-BPS states have $K>0$.
It follows from \levelmatch\ that if momentum and winding are nonzero a BPS state  can have a large $\bar N$, corresponding to a highly excited
string.  Since BPS states are exactly stable, in order to discuss the lifetime and decays of
states, we have to move away from the BPS bound. In the following we define a measure
of a state's closeness to BPS.

\eqn\nbsmeasure{n_1 = {2K\over m_0^2}, \quad n_2 =  {2(\bar N-1) \over m_0^2}}
A state is close to BPS if $n_1\to 0$ (which implies from \massshell, that there are few
left oscillators). For fixed $n_1$, the more excited a string is, the larger
$n_2$ (since from \massshell\ this implies the largest allowed number of right oscillators).  For example,
a state with large $l$, small $K$,  and  zero winding, $m=0$, will have $n_1\sim 0$
and is therefore near BPS. However, in this case, $k_L=k_R$ and so, $n_2 \sim  0$, meaning
$\bar N$ is small and the string is not excited. On the other hand, $l=m$
with large $l$ and small $K$ would yield a state both close to the BPS bound and highly excited.
It is this last type of state which most clearly exhibits the anticipated long lifetime.

\fig{Range of BPS and extremality measures $n_1$ and $n_2$}{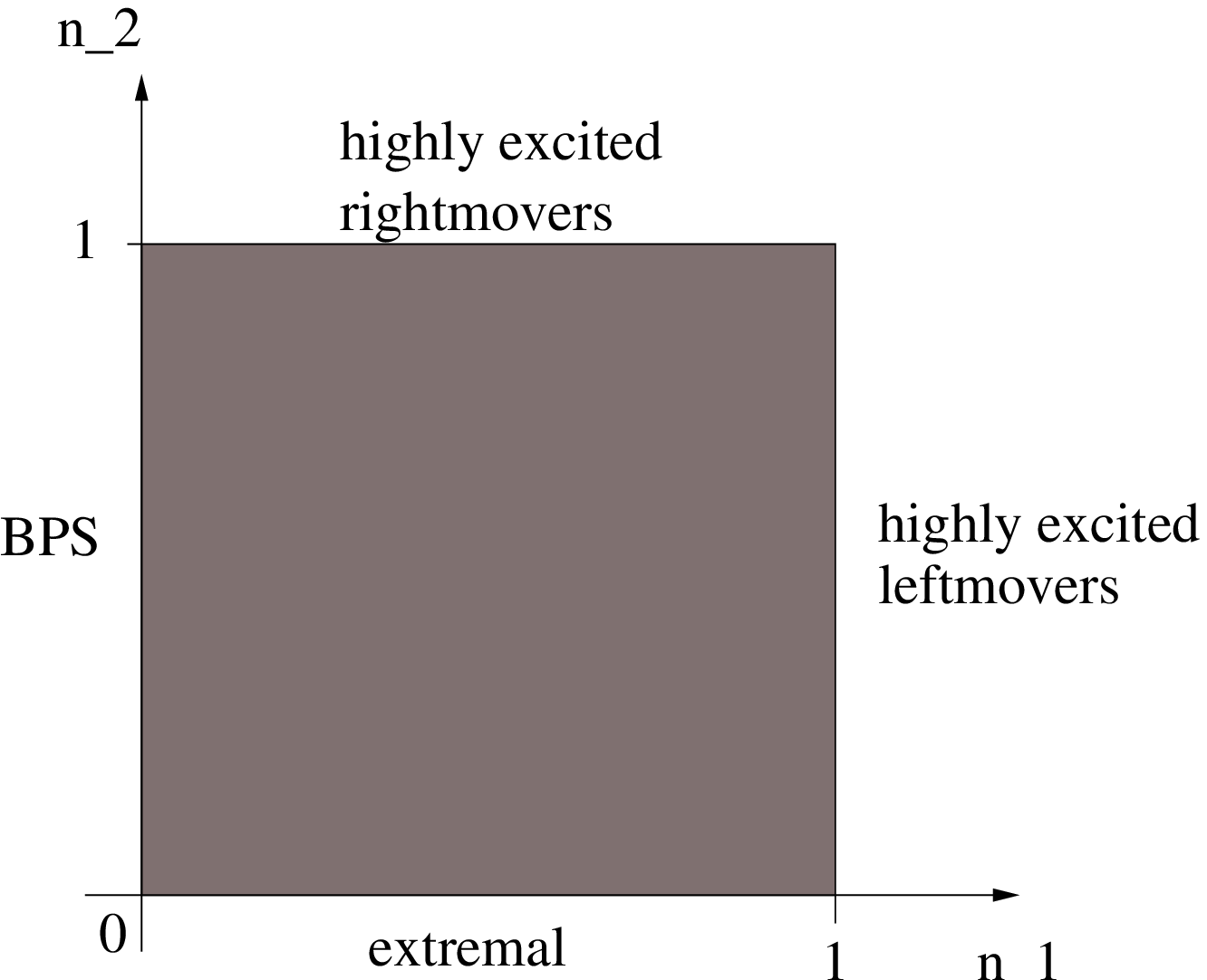}{2.5in}

A state which  saturates  the BPS bound  \BPSbound\ has equal mass and charge.
In the circle compactification there is a second $U(1)$ charge $k_R$ and hence,
there is another limit in which mass is equated with this charge, the so called 'extremal' limit
\eqn\extrcon{m_0^2= k_R^2}
The condition for a highly excited extremal state is that $n_2\to 0$ and $n_1\to 1$.
This has the same form as the BPS condition with the role of $n_1$ and $n_2$ exchanged.
Note that this bound is, however, unrelated to supersymmetry.

\subsec{States and vertex operators}

In order to calculate the decay rate and lifetimes one has to choose a particular state
and compute the related vertex operator. In  the lightcone gauge,
the massive spectrum is easy to construct \GreenSP\PolchinskiRQ;
solving the covariant physical state conditions for general massive states is quite
complicated. We consider special states and polarizations to simplify
the calculation. It would be interesting to generalize the calculations in this paper to more generic states.

The left moving part of the state is taken to be
\eqn\physta{\xi_{\mu\nu_1 \cdots \nu_{K}} b^\mu_{-{1\over 2}}
a_{-1}^{\nu_1}a_{-1}^{\nu_2} \cdots  a_{-1}^{\nu_{K  }} \mid 0 \rangle}
Physical state constraints of Virasoro currents and supersymmetric currents
$L_1\mid states\rangle=0$ and $G_{1/2}\mid state\rangle=0$ necessitate that $\xi_{\mu\nu_1
\cdots \nu_{K}}$ is transverse to the momentum in all indices.
$G_{1/2}\mid state\rangle=0$ also insures symmetry under any 2 index exchange.
Finally, $L_2\mid state\rangle=0$ and $G_{3/2}\mid state\rangle=0$ guarantee tracelessness in any pair of indices.
\eqn\phystab{\xi^{\nu_1}_{\;\;\nu_1 \nu_2 \cdots \nu_{K}}=0, \quad k^\mu \xi_{\mu\nu_1 \cdots \nu_{K}}=0 }
The polarization tensor is normalized as
$\xi_{\mu\nu_1\cdots\nu_{K}}\xi^{\mu\nu_1\cdots\nu_{K}}=1$.

The corresponding vertex operator in the $-1$ picture is  given
by
\eqn\phystb{V^{(-1)}_L= {1\over \sqrt{K!}} \xi_{\mu\nu_1 \cdots \nu_{K}} e^{-\phi}
\psi^{\mu}\partial X^{\nu_1}\partial X^{\nu_2}\cdots \partial X^{\nu_{K}}
 }
In a  one loop calculation  the vertex operator in
the zero picture instead of \phystb\ is needed. The picture changing
operator is given by
\eqn\piccha{\Xi(z) = e^{\phi} \partial X_\mu \psi^\mu(z)}
and the zero picture operator is given by
 \eqn\zerpic{V^{(0)}(w)= \lim_{z\to w} \Xi(z) V^{(-1)}(w)\mid _{(z-w)^0}}
i.e. one picks the term proportional to $(z-w)^0$ in the OPE of the
vertex operator in the $-1$ picture  and the picture changing operator.
The normalized vertex operator $V^{(0)}_L$ is then given by
\eqn\verlft{\eqalign {V^{(0)}_L =&{\sqrt{K}\over \sqrt{(K-1)!}} \xi_{\nu\nu_1 \cdots \nu_{K}}
\psi^{\nu}\partial\psi^{\nu_1}\partial X^{\nu_2}\partial X^{\nu_2}\cdots \partial X^{\nu_{K}}
+\cdots }}
where the dots denote terms which vanish in the two point function because of the sum over the spin structures.

The rightmoving part of the state is taken to be
\eqn\rmvtb{\bar \xi_{\rho_1 \cdots \rho_{\bar N} } \bar
a_{-1}^{\rho_1}\cdots \bar
a_{-1}^{\rho_{\bar N}} \mid k\rangle}
and  the associated normalized vertex operator is given by
\eqn\rtmva{V_{R}= {1\over \sqrt{\bar N ! }} \bar \xi_{\rho_1 \cdots \rho_{\bar N}} \bar \partial
X^{\rho_1} \cdots \bar \partial
X^{\rho_{\bar N}}  }
This  state satisfies the physical state conditions if the polarization
tensor $\bar \xi$ is
transverse, symmetric  and traceless. The polarization tensor is normalized as
$\bar \xi_{\nu_1\cdots \nu_{\bar N}}\bar \xi^{\nu_1\cdots \nu_{\bar N}}=1$.
Note that we, furthermore, simplify the calculation by imposing the conditions
 that the polarization tensors $\xi$ and $\bar \xi$ are orthogonal and do not have any leg in the $S_1$ direction.
\eqn\orthocon{\xi_{\nu_1\cdots \nu_{K+1}} \bar \xi^{\nu_1}_{\;\;\mu_2\cdots \mu_{\bar N} }=0, \quad \xi_{9\cdots \nu_{K+1}} = 0, \quad \bar \xi_{9\cdots \nu_{\bar N}}=0}
The complete vertex operator is then given by
\eqn\totvert{V(\xi, \bar \xi, k, k_L,k_R) = V_L^{(0)}(z) V_R(\bar
z)e^{i k_\mu X^\mu+ i
k_L X_L +i k_R X_R}}

\newsec{Calculation of the two point function}

The optical theorem relates the decay rate to the imaginary part of the mass
shift at one loop.
The mass shift is given by the one loop string amplitude with  two  vertex operators \totvert\ insertions.

\eqn\twoptama{\eqalign{\delta m^{2}=A_2&= g_s^2 \int d^2\tau \Big\langle V(0) \int d^2z
V^\dagger(z,\bar z)\Big\rangle }}
where the $z$ integral is over the usual fundamental torus domain,
i.e. the parallelogram stretched onto the unit vector along the real axis and the
$\tau=\tau_1+i\tau_2$ vector, and the $\tau$ integral is over the fundamental domain
of $SL(2,Z)$, the conformal transformations of the torus.

In order to calculate $A_2$ we use the vertex operator \totvert\ and evaluate the
various contractions.
\eqn\twoptamb{\eqalign{
 A_2 &=c' g_s^2  K(\xi\cdot\xi)(\bar\xi\cdot\bar\xi) \int {d^2\tau \, \tau_2}  \int d^2z
\langle e^{i k_{\mu} X^{\mu} }(0) e^{-ik_{\mu }X^{\mu}}(z,\bar z)\rangle  \langle e^{i k_L
X_L}(0) e^{-i k_L X_L}(z)\rangle \cr
&\quad \quad \langle e^{i k_R
X_R}(0) e^{-i k_R X_R}(\bar z)\rangle
 \langle \partial
X(0)\partial X(z)\rangle^{K-1}  \langle \bar \partial X(0)\bar
\partial X(\bar z)\rangle^{\bar N} \cr 
&\quad \quad  \sum_\nu \epsilon_\nu Z_\nu
\langle \psi \partial \psi(0) \psi\partial\psi (z)\rangle_\nu
Z_{het}(\bar \tau)Z,}}
where we used the fact that the vertex operators are normalized and the
orthogonality of the polarization tensor to eliminate right with left contractions. 
The constant $c'$ is an overall normalization which is independent of the initial state.
 Since we are interested only in the relative  comparison of rates and lifetimes we will set $c'g_s ^{2}=1$ when evaluating the amplitudes.
$Z_{\nu}$ is the partition function for fermions with spin structure $\nu$; $Z_{het}$ is the partition function for the bosons compactified on the $SO(32)$ lattice;
$Z$ is the partition function for bosons along the first ten dimensions and for the ghosts. The factor of $\tau_2$ is from gauge fixing the conformal symmetry of the torus
by choosing the first insertion point to be at 0.

The contractions and contributions to the partition functions are evaluated in Appendix A.  The fermionic contractions and the sum over spin structures is performed in (A.3), the SO(32) partition function is given in (A.7), the bosonic partition function is given in (A.8) and the bosonic  correlators are given in (A.11) and (A.14-15). Putting everything together the result is
\eqn\twoptamc{\eqalign{
A_2 & =c' g_s^2 K \sum_{n,w} \int
{d^2 \tau \over \tau_2^{9/2} }\int d^2 z   \sum_{i=1}^4 {\theta^{16}_i(0\mid \bar \tau)\over \eta^{24}(\bar
\tau) } \Big(
\partial_z^{\;2}\ln\big( \theta_1({z\over
2\pi}\mid \tau)\big) + {1\over 4 \pi \tau_2}\Big)^{K-1}  \Big(
{2\pi \theta_1 ({z\over 2\pi}\mid \tau) \over
\theta_1'(0\mid \tau)}\Big)^{2K} \cr
&\; \Big( \partial_{\bar z}^{\;2}\ln\big( \theta_1({\bar z\over
2\pi}\mid \bar \tau)\big) + {1\over 4 \pi \tau_2}\Big)^{\bar N} \Big(
{2\pi \theta_1 ({\bar z\over 2\pi}\mid \bar \tau) \over
\theta_1'(0\mid \bar \tau)}\Big)^{-2 + 2 \bar N}
e^{-{m_0^2 \over 2\pi \tau_2} z_2^2}  q^{{1\over 2}p_{L}^2}\bar q^{{1\over 2}p_{R}^2} e^{-i z p_{L }k_L } e^{i \bar z p_{R} k_R} }}

where $z=z_1+iz_2$. The sum over the integers $n,w$ represents the summation over the discrete loop momenta $p_{L}=n/R+ wR/2$ and $p_{R}=n/R-wR/2$ in the compactified direction,
which also makes them the left and right charges of one of the decay products.
The constant $c'$ is a normalization which is independent of all the parameters.
 The plan to evaluate the amplitude is as follows.
First, we expand the amplitude in infinite series in powers of $q$ and $e^{iz}$
and their complex conjugates.
Secondly, we perform an integration over the $\tau_1$ and $z_1$ which
yield constraints on the exponents of $q=e^{2\pi i \tau}$ and $e^{iz}$
and their complex conjugates.
Next, we cut off the infinite series to obtain finite polynomials,
by relating the exponents to the masses of the original state and its decay products,
and applying conservation of mass and charge. More precisely, not only is the series
cut off, but a finite list  of  the exact powers of $q$ and $e^{iz}$
and their complex conjugates which contribute to the amplitude is compiled.
The integrals for all the monomials, over $\tau_2$ and $z_2$
are, then,
performed analytically. Unfortunately, the final computation of coefficients of
all the monomials can only be performed,
for specified initial states,
using Mathematica for the power series
expansions of the expressions.

Using the formulae given in appendix B the holomorphic part of \twoptamc\ can be expanded as a
  power series in $q$ and $e^{iz}$
\eqn\pwsera{ \eqalign{&\Big(
\partial_z^{\;2}\ln\big( \theta_1({z\over
2\pi}\mid \tau)\big) + {1\over 4 \pi \tau_2}\Big)^{K-1}  \Big(
{2\pi \theta_1 ({z\over 2\pi}\mid \tau) \over
\theta_1'(0\mid \tau)}\Big)^{2K}\cr
&=\sum_{r=0}^{K-1} {1\over (4\pi \tau_2)^r} \pmatrix{K-1\cr r} \sum_{a,b}
C^{r,K}_{a,b} q^a (e^{iz})^b }}

The antiholomorphic part of the amplitude \twoptamc\ can be expanded as
 \eqn\pwserab{ \eqalign{&\Big(
\partial_{\bar z}^{\;2}\ln\big( \theta_1({\bar z\over
2\pi}\mid \bar \tau)\big) + {1\over 4 \pi \tau_2}\Big)^{\bar N}  \Big(
{2\pi \theta_1 ({\bar z\over 2\pi}\mid \bar \tau) \over
\theta_1'(0\mid \bar \tau)}\Big)^{2\bar N -2} \sum_{i=1}^4
{\theta^{16}_i(0\mid \bar \tau)\over \eta^{24}(\bar
\tau) } \cr
&=\sum_{s=0}^{\bar N} {1\over (4\pi \tau_2)^s} \pmatrix{\bar N \cr s}
\sum_{c,d}
\bar C^{s,{\bar N}}_{c,d} \bar q^c (e^{-i\bar z})^d }}
Using \pwsera\ and \pwserab\ the two point amplitude $A_2$ becomes
\eqn\twoptamd{\eqalign{A_{2} &= c' g_s^2 K\sum_{n,w} \int d^{2}\tau \int d^{2}z \sum_{r=0}^{K-1}
\sum_{s=0}^{\bar N}
{\tau_{2}^{-(r+s+{9\over 2})}\over (4\pi)^{r+s}}\pmatrix{K-1\cr r}\pmatrix{\bar N  \cr s}
e^{-{m_0^2 \over 2\pi \tau_2} \nu_2^2}  \cr & \times \sum_{a,b} \sum_{c,d} C_{a,b}^{K,r}
 \bar C_{c,d}^{\bar N,s} q^{a+{1\over 2} p_{L}^{2}}  \bar q^{c +{1\over 2} p_{R}^{2}}
 e^{iz( b - k_{L }p_{L} )} e^{-i\bar z( d+ k_{R} p_{R})}  }}

The integration over the modulus $\tau$  in \twoptamd\ is over the fundamental
domain of $SL(2,Z)$. However, as explained in the next subsection or in \WilkinsonTB,
if one is interested in the imaginary part of $A_{2}$ only (as we are), the
fundamental domain can be replaced by a regular strip $-1/2 <\tau_1<1/2$
along the entire length.

   The integration over $\tau_{1}$ and $z_{1}$ then take the form:
\eqn\taunurealintegrals{\int_{-1/2}^{1/2}d\tau_1e^{2\pi i\tau_1(nw+a-c)},
\int^{2\pi}_0 dz_1e^{i z_1(-nm-wl+b-d)}
}
so a non-zero amplitude requires the $a-c+nw=0$ and
$b-d-nm-wl=0$. After performing the integrations over $\nu_{1}$ and $\tau_{1}$ the
two point amplitude becomes
\eqn\twoptame{\eqalign{A_{2}&= 2\pi c' g_s^2 K\int d \tau_2 \int d \nu_2
\sum_{n,w}\sum_{r=0}^{K-1}\sum_{s=0}^{\bar N} \pmatrix{K-1\cr r}\pmatrix{\bar N  \cr s}
\sum_{a,b}  C_{a,b}^{K,r} \bar C_{a+nw,b-nm-wl}^{\bar N,s}  \cr
&{\tau_{2}^{-(r+s+{9\over 2})}\over (4\pi)^{r+s}} e^{-2\pi
 \tau_2 ( 2 a + (n/R+ wR/2)^2 )}
 e^{-  \nu_2 (2b - 2nl/R^2 - wm R^2/2-
nm-wl)} e^{-{m_0^2 \over 2 \pi \tau_2} \nu_2^2 }}}

\subsec{Calculation of decay rate}
After a change of variables $2\pi \tau_{2}=t$ and $\nu_{2}= t y$ the integrals in \twoptame\ are all  of the following form
\eqn\intega{\tilde a_{2} = \int_{0}^{\infty} dt \int_{0}^{1} dy  \; t^{\alpha -1} e^{- t \big( m_{0}^{2} y^{2} -y (m_{0}^{2} -m_{1}^{2}+m_{2}^{2})  + m_{2}^{2}\big)}}
where $\alpha=-r-s-5/2$ and is negative.

The choice of variable names $m_1$, $m_2$ is not accidental. By considering a Schwinger parametrization of a Feynman integral for a loop amplitude, we can  convince ourselves that the coefficients
of powers of $y$ are  associated with the masses of the decay products as above \MitchellQE,\WilkinsonTB,\ChialvaHG.
For completeness, the argument is reproduced in Appendix C.

A comparison of \intega\ with    \twoptame\ gives the following expressions for the masses $m_{0},m_{1},m_{2}$
\eqn\abcdefb{\eqalign{m_0^2&=\Big({l\over R}+ {mR\over 2}\Big)^2  +2 K\cr
m_1^2 &=  \Big( {l-n\over R}+ {(m-w)R\over 2}\Big)^2 +2(a+b+K) \cr
m_2^2&=  \Big({n\over R}+ {wR\over 2}\Big)^2 +2a} }

Note that a comparison with the mass-shell  condition \massshell\  leads to an identification of  the momentum $k_{1,L}, k_{2,L}$ as well as the oscillator level $K_{1},K_{2}$ of the two decay products with mass $m_{1}$ and $m_{2}$ respectively.
\eqn\identa{\eqalign{k_{1,L} &=  k_L-p_L=\Big( {l-n\over R}+ {(m-w)R\over 2}\Big), \quad K_{1}= a+b +K\cr
k_{2,L} &=p_L=  \Big( {n\over R}+ {wR\over 2}\Big), \quad K_{2}= a  }}
where $k_L$ is the leftmoving compact momentum of the initial state and $p_L$ is the compact loop momentum.
As we shall see later the power series expansion of the amplitude limits the range of b to be
$-a-K \leq b \leq -a$.

The optical theorem relates the imaginary part of the loop amplitude to the summed squares
of the tree three point amplitudes. The loop amplitude is formally real
but the imaginary part comes from the  analytic
continuation of the amplitude into
a region where it is divergent. From \intega\ we can see that the integral is divergent when the
polynomial coefficient of t in the exponent,
\eqn\inergb{P(y)= m_{0}^{2} y^{2} -y (m_{0}^{2} -m_{1}^{2}+m_{2}^{2})  + m_{2}^{2}}
is negative, which occurs between the roots of this polynomial,
 \eqn\loopi{y_\pm ={m_0^2-m_1^2+m_2^2\over 2m_0^2}\pm {1\over m_0}\sqrt{{(m_0^2-m_1^2+m_2^2)^2\over 4m_0^2}-m_2^2}}

 This restricts the domain of integration in the $y$ variable. Note that the roots are
 real and the imaginary part of $A_2$ exists if and only if $m_0>m_1+m_2$.

 Let us now turn to the analytic continuation of the  $t$ integral.
The two divergences to deal with are a pole at $t=0$ and an
essential singularity at $t=\infty$. We will analytically continue in $P$ and $\alpha$.
 The integral is
well defined for $Re[P]>0$, $Re[\alpha] \geq 1$. We make the substitution $u=tP$. The $t$ integral is then:
\eqn\analcontone{
{1 \over P^\alpha}
\int_{P\epsilon}^{\infty} \exp^{-u}u^{\alpha-1}du={1\over P^\alpha}
\int_{0}^{\infty}\exp^{-u}u^{\alpha-1}du-{1\over P^\alpha}\int_{0}^{P\epsilon}\exp^{-u}u^{\alpha-1}du
}
$P\epsilon$ is the variable lower limit inherited from the
 $\tau$ domain.
In a moment we will see that the exact lower limit does not matter in the calculation,
because the second
integral is real for all values of $P$ and $\alpha$, and therefore, the complicated
 $\tau$ domain can be
replaced by a strip extending down to $0$. (It may worry the reader that this integral
domain now includes infinite copies of the fundamental domain and should diverge by modular
invariance. However, modular invariance has, already, been broken in our calculation because
we do not keep the entire modular invariant expression for the amplitude, but instead pick
out only a finite number of monomials.)
More explicitly, if we analytically continue
 the first integral
to negative $P$ we obtain:

\eqn\analcontint{ {1\over (-1)^\alpha(-P)^\alpha}
\int_{0}^{\infty}\exp^{-u}u^{\alpha-1}du={\pm i\over (-P)^\alpha}
\Gamma(\alpha)={\pm i \over (-P)^\alpha}
{\pi \over \Gamma(1-\alpha) \sin(\alpha\pi)}}
where we used that $\alpha$ is a half-integer (see \twoptame).\foot{Note that, were the number
of compactified dimensions even, $\alpha$ would be an integer and the analytic continuation
 would need slight modification.}  The sign in \analcontint\ is chosen to give a positive mass shift. Now the expression can be analytically continued for
 negative $P$ and negative $\alpha$. For such values the expression is pure imaginary.
Now let's see that there's no imaginary contribution from the second integral in
\analcontone\ . Let $P$ take its negative value and make the variable change :
$u\rightarrow-u$ (or equivalently make the $u=tP$ substitution on the $0-\epsilon$
integral before doing any analytic continuation in $P$)

\eqn\analcontintreal{(-1)^{\alpha-1}{1\over (-P)^\alpha}\int_{0}^{-P\epsilon}\exp^{u}u^{\alpha-1}du
=-{1\over (-P)^\alpha}\int_{0}^{-P\epsilon}\sum_{n} {u^n \over n! u^{\alpha-1}}
=\epsilon^\alpha\sum_n {(-P\epsilon)^n\over n!(n-\alpha)}}

This expression is convergent with no poles and it is still so if analytically continued
for negative $\alpha$. Since this integral has no imaginary part and the first integral
had no dependence on the lower bound, the fundamental domain of $\tau$ can be replaced by
a strip extending all the way down to zero. All that's left is the $y$ integration.
\eqn\yintegral{\int_{y-}^{y+}(-P)^{-\alpha} dy=\int m_0^{-2\alpha}(y-y_-)^{-\alpha}(y_+ - y)^{-\alpha}m_0^{-2\alpha}(y_+ - y_-)^{1-2\alpha}{{\Gamma(1-\alpha)^2} \over {\Gamma(2 -2\alpha)}}}

where the roots, $y_+$ and $y_-$, were defined in \loopi\ and $\alpha=-r-s-5/2$.

\medskip
Combining with the $t$ integral we obtain
\eqn\loopl{Im(\tilde a_2)= \Theta(m_{0}-m_{1}-m_{2}) {2^{2r+2s+6}\pi \over m_0} {\Gamma(r+s+7/2)\over
\Gamma(2r+2s+7)} \Big( {(m_0^2-m_1^2+m_2^2)^2\over 4m_0^2}-
m_2^2\Big)^{r+s+3}}
Where the $\Theta$-function enforces the on-shell condition $m_{0}> m_{1}+m_{2}$.
Applying \loopl\ to the terms appearing in the summation \twoptame\ gives the imaginary
part of the two point amplitude

\eqn\twpima{\eqalign{Im(A_2) &= c' g_s^2 K\sum_{r=0}^K \sum_{s=0}^{\bar N} \sum_{n,w}
\sum_{a,b} \Theta(m_0-m_1-m_2) \pmatrix{ K-1\cr r}
\pmatrix{\bar N\cr s}\pi^{11\over 2}  2^{21/2+r+s} \cr
 \times&  C_{a,b}^{r,K} \bar C_{a+nw, b-nm-
wl}^{s,\bar N} {\Gamma({7\over 2}+ r +s) \over \Gamma(7 + 2 r +2 s)}
{1\over m_0}  \Big( {{(m_0^2- m_1^2+m_2^2)^2}\over{4m_0^2}} -
m_2^2\Big)^{3+ r+s} }}
The total decay rate is given by
\eqn\decz{\Gamma = {Im(\delta m^{2})\over 2 m_{0}}={Im(A_{2})\over 2m_{0} }}
The lifetime of the unstable string state is the inverse of the total decay rate
\eqn\totaltb{T={1\over \Gamma}}
Note that the summation over the loop momentum $n$  and winding $w$ and $a,b$  sums over all particles with mass $m_{1}$ and $m_{2}$ given by \abcdefb. If the on shell condition is fullfilled and  $C_{a,b}^{r,K} \bar C_{a+nw, b-nm-
wl}^{s,\bar N} $ is nonzero then the the original string state can decay in this channel.
The decay rate in this channel is given by the imaginary part \twpima\ for a fixed $n,w,a,b$.

\newsec{Decays of near BPS states}
Massive BPS states are generically stable against decays and hence one has to move away
from exact BPS states to investigate the lifetime of excited string states.
An excited string state represented by the vertex operator \totvert\ is characterized
by the momentum, $l$, winding number, $m$, and the left and right moving excitation
level $K$ and $\bar N$, respectively. The level matching condition \levelmatch\ provides one relation
 among the four parameters. The measure of how close a state is to BPS is therefore not
  unique and depends on the definition. In the following we will consider states which
  have $n_{1}$ close to 0 and $n_{2}$ close to 1. This can be achieved by choosing $K$
  to be close to zero and choosing $l,m$ such that  $k_{L}$ is large and $k_{R}$ is
  small, so that the state has a large $\bar N$. In the following, we also fix the
  radius of the compactification circle to be $R=1$.

It is expected that the states which are closest to BPS are the longest lived for a
given mass. It is therefore interesting to determine the dependence of the lifetime on
the mass $m_{0}$ of the initial state. In the following plot, states with $K=1$ and
$p_{R}=0,{1\over 2}$,$1$ were considered with $p_{L}$ varying such that the
mass ranges from $m_{0}^{2}=98$ to $m_{0}^{2}=308$.

\fig{Log/Log plot of lifetime $T$ of $K=1$ as a function of  mass $m_{0}^{2}$.
The masses range from }{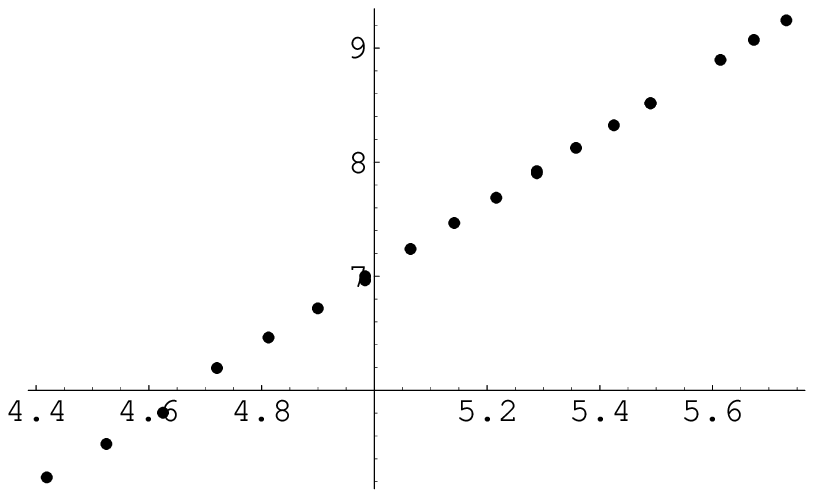}{2.5in}

A $\chi^{2}$-fit for the mass dependence of the lifetime gives
\eqn\lifet{T= e^{-8.11} (m_{0})^{6.06} }
As we shall show below, the exact behavior of the lifetime of a near BPS state  in the limit of
large masses is $T\sim m_{0}^{6}$.

\bigskip
In the following table we list the decay rates and the dominant decay channel of near BPS states with momentum $l=9$ and winding number $m=16$ for varying $K$. The lifetimes are calculated using \decz\ setting $g_{s}^{2}c'=1$ to fix the normalization.
\bigskip

\begintable
K|  $m_{0}^{2}$| $ n_{1}$,|$ n_{2}$| lifetime |  dominant decay channel |  per cent\elt
1|291|0.0068|0.996|8705.5|  $K_{1}=0,m_{1}^{2}=289, m_{2}^{2}=0$|99.9\elt
2|293|0.013|0.996|4376.2|  $K_{1}=1,m_{1}^{2}=290.7, m_{2}^{2}=0$|98.3\elt
4|297|0.0269|0.996|2212.5|  $K_{1}=3,m_{1}^{2}=294.8, m_{2}^{2}=0$|95.0\elt
6|301|0.039|0.996|1492.1|  $K_{1}=5,m_{1}^{2}=298.9, m_{2}^{2}=0$|92.0
\endtable
\medskip
\centerline{{\bf Table 1}: Lifetime and dominant decay channel of near BPS states}
For near BPS states the lifetime increases when $n_{1}$ gets closer to zero, which is
a measure of closeness to BPS, hence the closer to BPS a state (of comparable mass) is,
the longer lived it is.  The decay is dominated by a single decay channel, where one
of the decay products is a massless graviton and the other is a massive string state
where $K$ has been decreased by one (subleading decay channels decrease $K$ by more
than one unit but give a small contribution to the total rate).  Hence a massive near
 BPS state will gradually become more BPS by massless radiation.

The mass dependence \lifet\ and the dominance of the massless decay channel can be
understood in more detail by analyzing \twpima.
Since the decay amplitude is symmetric under interchange of $m_{1}$ and $m_{2}$,
 for a  massless  decay we can pick the second particle to be massless, i.e. $m_{2}=0$.
  Conservation of energy implies that  the first decay product has mass
  $m_{1}^2= k_{L}^{2}+2(K-\Delta K)$, where $\Delta K$ is a positive integer related
  to the decrease in non BPS-ness. For the massless decays the  last factor in \twpima\
   then becomes
\eqn\factorone{ \Big( {{(m_0^2- m_1^2+m_2^2)^2}\over{4m_0^2}} -
m_2^2\Big)^{3+ r+s} = \Big( {\Delta K^{2} \over m_{0}^{2} }\Big) ^{{3 +r+s}}}
The inverse powers of $m_0$ in \factorone\ suppress terms with $r,s>0$, unless these
inverse powers are balanced by combinatorial factors coming from the power series
expansion. The combinatorial factors on the holomorphic side are small for near BPS 
decays because they are of order $K$ and $K$ is small, but on the antiholomorphic side combinatorial factors can be large
since they can be of order $\bar N$, i.e. of order $m_0^2$ for near BPS states, so we can set $r=0$ to a good approximation,
but different $s$ can contribute equally.
Now let's see what powers of $q$, $\bar q$, $e^{iz}$, $e^{-iz}$ we're after.
It follows from  $m_2=0$ and \abcdefb\ that $a=0$ as well as $n=w=0$. This means that in
the power series expansion \twoptamd\  we are looking for the terms with $q^0$. In this
case the mass of the first decay product is given by $m_1^2=m_0^2 + 2b$ and it follows
from $m_0>m_2$  that we are looking for a negative power of $(e^{iz})^b$ in \pwsera.
From Appendix B we can see that if we keep only terms proportional to $q^0$ and set $r=0$ we get
\eqn\holoexpand{(\sum_{l=1}^{\infty}le^{izl})^{K-1}(ie^{-iz/2}(1-e^{iz}))^{2K}}
We can see that only the first negative power of $e^{iz}$ is present so $b=-1$. Comparing with \identa\
one sees, for $a=0$, $b=\Delta K$, which explains why the dominant decay channel has $\Delta K=1$.
On the antiholomorphic side, from \taunurealintegrals, we know that $c=0$ and $d=-1$, so we are looking for
$\bar q^0 e^{i\bar z}$. Keeping only $q^0$ terms from \pwserab\ we obtain
\eqn\antiholoexpand{\eqalign{({1 \over \bar q}+24)(&2+960\bar q)(1+(-6+2\bar N)\bar q)
(ie^{i \bar z /2}(1-e^{-i\bar z}-\bar qe^{-2i\bar z }))^{2\bar N-2} \cr
&\times\big( \big(\sum_{l=1}^{\infty}le^{-i\bar z l}\big)^{\bar N-s}
+(\bar N-s)\bar q(e^{-i\bar z}+e^{i\bar z})\big( \sum_{l=1}^{\infty}le^{-i\bar z l} \big)^{\bar N-s-1}\big)}}
Expanding the above expression shows that for small $s$ the coefficient of $\bar q^0 e^{i\bar z}$ goes like $\bar N^s$
so small $s$ contribute to the computation at the same order of $m_0$. (For large $s$ the combinatorial factors
no longer go like $\bar N^s$.) Therefore we see that, in the limit of large initial
mass, the decay rate of near BPS states behaves like $\Gamma\sim 1/m_0^6$ in accordance with the numerical result \lifet.

\newsec{Decays of other states}
The calculation of the decay rate and lifetimes is not limited to near BPS states.
It is an important question, whether the near BPS states are indeed the longest lived
states. To this end, we have calculated the lifetimes for all inital states of a given
fixed mass, $m_{0}^{2}=171$. For radius $R=1$, there are 134 possible initial states. In the following table, we list the ten longest lived states among them.

\medskip

\begintable
$k_{L}$| $ k_{R}$| $K$| $n_{1}$|$ n_{2} $ |  lifetime |  decay channel \elt
-13|-1|1|0.0117|0.9942| 1750|massless\elt
-13|1|1|0.0117|0.9942|1683|massless\elt
-13|-3|1|0.0117|0.9474|1645|massless\elt
-13|3|1|0.0117|0.9474|1197|massless\elt
-13|5|1|0.0117|0.8538|375.8|light\elt
-13|-5|1|0.0117|0.8538|99.75| light\elt
-13|7|1|0.0117|0.7135|85.95|light\elt
-5|13|73|0.8538|0.0117|62.24|light\elt
-7|13|61|0.7135|0.0117|41.95|light\elt
-11|-3|25|0.2924|0.9474|38.32|massless\elt
$\cdots$|$\cdots$|$\cdots$|$\cdots$|$\cdots$|$\cdots$|$\cdots$\elt 
 7|+13|61|0.7135|0.0117|0.0152|light\elt
-13|+13|1|0.0117|0.0117| 0.0111|massless
\endtable

\medskip
\centerline{{\bf Table 2}: Longest lived states of mass $m_{0}^{2}=171$}
\medskip
The longest lived states are indeed the near BPS ones with massless radiation as dominant decay channel.
A near extremal state appears as the 8th longest lived state, the lifetime is however 30 times shorter than
that of the logest lived near BPS state. The dominant 'light' decay channels consist of BPS states with nonzero
winding or momentum and a mass close to zero. The states with dominant 'light' decay channels
exhibit shorter lifetimes compared to the states with dominant massless decays.

For comparison purposes, we have listed the shortest lived state among the 134 states of mass
 $m_{0}^{2}=171$. Note that the state is technically near BPS since $n_{1}$ is small, however,
  it is not highly excited, since $n_{2}$ is small. Hence, it is the highly excited near BPS states
  which are long lived.

In contrast to the BPS condition, extremality \extrcon, is not related to supersymmetry and it is an interesting question whether highly excited extremal states are also long lived. Already from table 2 it is clear that extremal states are shorter lived than BPS states (for a particular mass). We have calculated the lifetimes of near extremal states with $\bar N=2$ and masses ranging from $m_0^2=38$ to $m_0^2=258$.

\fig{Log/Log plot of lifetime $T$ of $\bar N=2$ near extremal states  as a function of  mass $m_{0}^{2}$.}{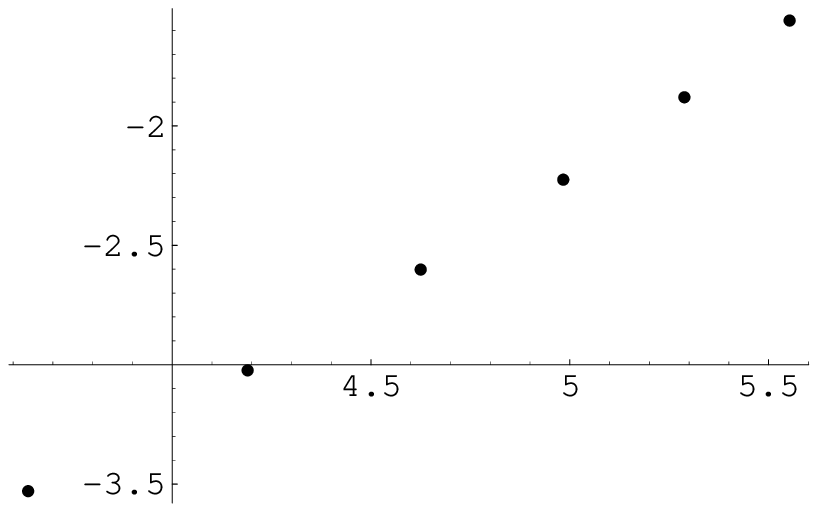}{2.5in}

A fit of the data produces a  lifetime growing as $T\sim m_0^{2.04}$, however, note that, compared to the near BPS states, the lifetimes of the near extremal states are several orders of magnitude shorter.

A general  feature of all decays, whether  the initial state is near BPS or not, is that all decays are dominated
(to at least $99\% $ of the amplitude) by channels where one decay product (say the first one) has oscillator level
$K_{1}=0$. Note that, in general, the dominant decay channel may be not massless, but the mass
 of the state will come from momentum and/or winding, not from left oscillators.
To understand this, we must again look at the expansion of the holomorphic piece \pwsera\ , but we do not assume $a$ is 0.
Remember that $a=K_2$ and $a+b+K=K_1$, which imply together with mass conservation, that $-K\leq a+b < 0$ and $a<K$,
where $a$ is the power of $q$ and $b$ is the power of $e^{iz}$. Also, remember that terms with $r>0$ are suppressed.
One obvious term one can pick out for $r=0$ is $(qe^{-iz})^{K-1}e^{-izk}$. (This is obtained by picking up from (B.2)
only terms from second sum with $l,n=1$ and only terms with $n=1$ in (B.3) .) This term has $a+b=-K$, i.e. $K_1=0$,
and $a=K-1$, as we wanted to argue. Now suppose we try to pick $a+b$ less negative.
Whether you do this by picking $n>0$ in one of the $2K$ factors in (B.3) or by choosing $n>1$ in one factor of (B.2),
the power of $q$ becomes too large and we are forced to use $r>0$, which produces suppressed decay rates.


\newsec{Discussion}
In this paper we have analyzed the decay of highly excited string states for compactified heterotic strings.
 We found that the longest lived states are highly excited, near BPS states, for which the lifetime grows as
  $T\sim 1/g_s^2  (m_0)^6$ for large masses. The decay of such states is dominated by massless radiation, which reduces the
  non-BPS'ness $K$ by one. The end result of such a  decay (over several steps) is a stable BPS state. For large mass
  $m_0$, the near BPS states is very long lived.   Near extremal states and generic states, which have both left and
  right excitations, are much shorter lived, although the lifetime of near extremal states grows with mass.  It is an
  interesting question,
   whether the decay of excited strings can be used for exploring the string/black hole correspondence. To achieve this,
    one has to consider more generic string states, instead of states on the leading Regge trajectories.
    Some steps were taken in this direction, for the comparison of absorption of massless scalars by perturbative strings
    and BPS black holes, in a recent paper \CornalbaHC, and it would be interesting to generalize this work to decays of near BPS states.
     We plan to address this question in the future.

\bigskip

\noindent {\bf Acknowledgments}

This work is supported in part by NSF grant PHY-0456200 and NSF grant PHY-0245096. We are grateful to Eric D'Hoker, Per Kraus and Iosif Bena for useful conversations.

 \appendix{A}{Collection of formulae for one loop string amplitudes}

The partition functions of the fermionic leftmovers with
even spin structure $\nu=2,3,4$  are
\eqn\spinstra{Z_2= {\theta_2(0\mid \tau)^{4}\over \eta(\tau)^{4}}, \quad Z_3= {\theta_3(0\mid \tau)^{4}\over \eta(\tau)^{4}}, \quad  Z_4= {\theta_4(0\mid \tau)^4\over \eta(\tau)^{4}} }

The propagator  for two world sheet spinors for the three even  spin structures, denoted by
$\nu=2,3,4$, is given by
\eqn\spincla{\langle
\psi^\mu(z)\psi^\nu(w)\rangle_\nu=\eta^{\mu\nu}{1\over 2\pi} {\theta_\nu( {z-w\over 2\pi}\mid \tau)\theta'_1(0\mid \tau)\over
\theta_\nu(0\mid \tau)\theta_1({z-w\over 2\pi}\mid \tau)}}

Correlation functions for the odd spin structure $\nu=1$  vanish, unless there
are ten fermionic zero modes inserted, which is not the case for our amplitude.

After summing over
spin structures, the correlators involving fewer than four fermionic fields
vanish.
Also, fermionic correlators involving four fermions without derivatives vanish.
The only nonzero fermionic correlator in our calculation
of the decay rate is
\eqn\fermcora{\sum_\nu \epsilon_\nu Z_\nu(\tau)\langle \psi^\mu(z)
\partial_z \psi^\sigma(z)  \psi^\rho(w) \partial_w \psi^\lambda(w)\rangle_\nu
=(\eta^{\mu\lambda}\eta^{\sigma\rho}+\eta^{\mu\rho} \eta^{\sigma\lambda})\eta(\tau)^8}

where  $\epsilon_\nu$ are  phases implementing the
GSO projection via $\epsilon_2=-1, \epsilon_3= 1, \epsilon_4=-1$.

Calculating \fermcora\ and vanquishing the other fermionic correlators is accomplished by applying the following
Riemann identity to derivatives of \spincla\ and \spinstra\ .
\eqn\riemanid{{1\over 2} \sum_\nu \epsilon_\nu \prod_{i=1}^4\theta_\nu(v_i\mid \tau) = -\prod_{i=1}^4 \theta_1(v_i')}
where
\eqn\riemanidb{\eqalign{v_1'&={1\over 2} (-v_1+v_2+v_3+v_4),  \quad v_2'={1\over 2} (+v_1-v_2+v_3+v_4)\cr
v_3'&={1\over 2} (v_1+v_2-v_3+v_4), \quad v_4'={1\over 2} (v_1+v_2+v_3-v_4)}}
and
\eqn\rimanidc{\theta_1'(0\mid \tau)=2\pi \eta(\tau)^3}

The partition function for the bosonic rightmovers in the $SO(32)$ lattice directions is
\eqn\partitionright{Z_{het}={1\over 2}\sum_{i=1}^{4}{\theta_i(0 \mid \bar\tau)^{16} \over \eta(\bar \tau)^{16}}}

The partition function (including the zero modes integral) of the nine left and right moving bosons
in the uncompactified spacetime directions is
\eqn\partitionboson{Z_{st}=\int d^9 p \, e^{-p^2\pi\tau_2} {1\over \eta(\tau)^9\eta(\bar\tau)^9} ={1\over \tau_2^{9/2} \eta(\tau)^9\eta(\bar\tau)^9}}

The correlation function for noncompact worldsheet bosons
 $X^\mu ,
\mu=0,\cdots,8$ is given by
\eqn\xxcor{\langle X^\mu(z,\bar z)
X^\nu(w,\bar w)\rangle= -{\alpha'\over 2} \eta^{\mu\nu}\ln \Big(
\mid  E(z,w)\mid ^2 + {1\over 8 \pi \tau_2}( z-w-\bar z+\bar w)^2\Big)}
where the prime form $E$ is defined as
\eqn\primef{E(z,w)=2\pi {\theta_1({z-w\over 2\pi}\mid \tau)\over \theta'_1(0\mid \tau)}}

The correlators involving $\partial X^\mu$ can be derived from \xxcor\ and read
\eqn\dxdxcor{\eqalign{\langle \partial X^\mu (z) \partial X^\nu (w)\rangle = &\eta^{\mu\nu}  {\alpha'\over 2} \partial_z^{\;2}\ln\Big( \theta_1({z-w\over
2\pi}\mid \tau)\Big) +\eta^{\mu\nu} {\alpha'\over 8 \pi \tau_2} \cr
\langle \bar \partial X^\mu (\bar z) \bar \partial X^\nu (\bar w)\rangle =  &\eta^{\mu\nu}{\alpha'\over 2} \partial_{\bar z}^{\;2}\ln\Big(
\theta_1({\bar z-\bar w\over
2\pi}\mid \bar \tau)\Big) +\eta^{\mu\nu} {\alpha'\over 8 \pi \tau_2}
}}

Using \xxcor\ the following correlation function is also calculated:
\eqn\verlc{\langle \prod_{i=1}^n e^{i k_i X(z_i,\bar z_i)}\rangle= \prod_{i<j}\mid  E(z_i,z_j)\mid ^{\alpha' k_ik_j} \exp\big( {\alpha'
\over 16 \pi \tau_2 }\sum_{i<j} k_i k_j (z_i-\bar z_i -z_j+\bar
z_j)^2\big)}

In the 9th direction, compactified along the $S_1$, the zero modes integral becomes a sum since the momenta are discrete,
so the partition function for the boson along the $S_1$ direction is
\eqn\partitioncompact{Z_{S_1}=\sum_p e^{i\pi\tau p^2-i\pi\bar\tau\bar p^2}{1\over\eta(\tau)\eta(\bar\tau) }}

For compact bosons we need a 'chiral' version of \verlc (derived in
\VerlindeKW),
\eqn\verl{\langle \prod e^{i k X(z_i)} \rangle = \exp( i p \sum_{i=1}^n k_i z_i)\prod_{i<j} E(z_i,z_j)^{{\alpha'\over 2}  k_i k_j}}
where $p$ is the loop momentum and $k_i$ is the left moving incoming momentum.

For the
antiholomorphic  correlators one gets
\eqn\verlb{\langle \prod e^{i \bar k_i \bar X(\bar z_i)} \rangle = \exp(
-  i \bar p \sum_{i=1}^n \bar k_i \bar z_i)\prod_{i<j}
\bar E(\bar z_i,\bar z_j)^{{\alpha'\over 2}\bar k_i \bar k_j}}

where $\bar p_i$, $\bar k_i$ are the corresponding right moving momenta. Note that, since there is winding,
a compact boson has $k_i\neq\bar k_i$.
We can check, that if multiply \verlb\ , \verl\ , and $Z_{S_1}$ and replace the sums over $p_i$ by integrals, we recover
the nonchiral correlator times the partition function of one uncompactified boson  i.e. \verlc\ multiplied by
${1\over \tau_2^{1/2}\eta(\tau)\eta(\bar\tau)}$.

The last piece we need is the ghost partition function
\eqn\zghost{Z_{gh}={\eta(\tau)^2\eta(\bar\tau)^2 \over \tau_2}}
Finally, to account for the gauge
fixing of the conformal invariance of the torus i.e. fixing the first insertion point to be at $0$,
we must augment $Z_{gh}$ with a factor of $\tau_2$.
All the factors of $\tau_2$ combine to yield $\tau_2^{-9/2}$.


For notational convenience we define a single partition function combining the bosonic and ghost partition function
\eqn\bigz{Z=Z_{st}Z_{S_1}Z_{gh}}

\appendix{B}{Expansion of theta functions}
In this appendix we gather the formulae for the series expansion of theta functions and related objects in $q$ and $e^{iz}$. The basic expansion formulae for the theta functions are
\eqn\thetaa{\eqalign{\theta_1(v\mid \tau)&=i \sum_n(-1)^{n}
 q^{{1\over 2}(n-{1\over2} )^2}  e^{iz (n-{1\over 2})}  \cr
 \theta_2(v\mid \tau)&=\sum_n
 q^{{1\over 2}(n-{1\over2} )^2}  e^{iz (n-{1\over 2})}  \cr
 \theta_3(v\mid \tau)&=\sum_n
 q^{{1\over 2}n^2}  e^{iz n}  \cr
 \theta_4(v\mid \tau)&=\sum_n(-1)^{n}
 q^{{1\over 2}n^2}  e^{iz n}  \cr
 }}
From these basic formulas the expansion of the terms appearing in appendix A can be derived
\eqn\expaa{\partial_z^{\;2}\ln\big( \theta_1({z\over
2\pi}\mid \tau)\big) = \sum_{l=1}^\infty l (e^{iz})^ l+ \sum_{l>0,n>0} l q^{nl}
\big((e^{iz})^l + (e^{iz})^{-l}\big)}
and
\eqn\expab{{2\pi \theta_1 ({z\over 2\pi}\mid \tau) \over
\theta_1'(0\mid \tau)} =i e^{-iz/2} \sum_{n\in Z} (-1)^{n} q^{{1\over 2}(n(n-1) )}  e^{iz n}  {1\over \prod_{m>0} (1-q^m)^3}}
and similarly for the antiholomorphic terms.

 \appendix{C}{Identification of Masses in Amplitude by Comparison with Schwinger Reparametrized Feynman Integral}
Consider the Feynman integral for the one loop amplitude. For simplicity, let's first consider the loop particles
to be scalars.

\eqn\feynman{
\int_{-\infty}^{\infty}d^d p {1 \over p^2+m_2^2}{1 \over (p-k)^2+m_1^2}}
where $p$ is the loop momentum, $k$ is the incoming momentum, $d$ is the uncompactified spacetime dimension
(which is 9 for us).
We preform a Schwinger reparametrization and obtain

\eqn\schwinger{
-\int_{-\infty}^{\infty}d^d p\int_{0}^{\infty}d\gamma\int_{0}^{\infty}d\beta \, e^{-i\gamma(p^2+m_2^2)}e^{-i\beta((k-p)^2+m_1^2)}}
Performing the momentum integration we obtain
\eqn\intoutmom{-\int_{0}^{\infty}d\gamma\int_{0}^{\infty}d\beta \,e^{-i\gamma m_2^2-i\beta(k^2+m_1^2)}e^{{-k^2\beta^2 \over i\gamma+i\beta}}\big({\pi \over i\gamma + i \beta}   \big)^{d \over 2}}

Finally let's change to variables $y={\beta \over \gamma +\beta}$, $t=i\gamma+i\beta$,
and rotate the $t$ contour to run along the real axis.
\eqn\stringvarint{\int_{0}^{\infty}dt\int_{0}^{1}dy \, \big({\pi \over t} \big)^{d-1\over 2}e^{-t(m_0^2 y^2- y(m_0^2+m_2^2-m_1^2)+m_2^2)}}
where we used $-k^2=m_0^2$ and that the jacobian for the variable change is $-t$.
This integral looks quite similar to \intega, except the power of $t$ in \intega is $-r-s-(d-1)/2$ whereas here it is
just $(d-1)/2$. The second issue is that there is no reason to assume that the decay products are scalars. Let us, then, consider what
happens with some arbitrary polarization. Note that the precise polarizations allowed by string theory are not arbitrary.
Furthermore, the states we are working with have even more specialized polarizations. We are not trying to establish an exact
correspondence between the string theory calculation and the equivalent field theory calculation, but only to extract the
 masses of the decay products.
To that end, we just note, that the integral must be a linear combination of integrals such as the following

\eqn\feynmanpolarized{
\int_{-\infty}^{\infty}{p^{2l}(pk)^c \over p^2-m_1^2}{1 \over (p-k)^2-m_2^2}d^d p}
The powers of $p$ must be expanded in components so the expression after Schwinger reparametrization looks something
like
\eqn\schwingerpolarized{
-\int_{-\infty}^{\infty}d^d p\int_{0}^{\infty}d\gamma\int_{0}^{\infty}d\beta \,
\sum_{ \sum j_i=2l+c} \prod_i D_i p_i^{j_i} e^{-i\gamma(p^2-m_2^2)}e^{-i\beta((k-p)^2-m_2^2)}}

where $D_i$ are some coefficients, which might depend on $k_i$. The thing to notice here is, that though this expression
is much messier than the one for scalars, the exponentials are unchanged,
so performing the gaussian integrals we get something like
\eqn\intoutmompolarized{
\sum_\sigma D'_\sigma \int_{0}^{\infty}d\gamma\int_{0}^{\infty}d\beta \,
e^{-i\gamma m_1^2-i\beta(k^2+m_2^2)}e^{{-k^2\beta^2 \over i\gamma+i\beta}}
\big({\pi \over i\gamma + i \beta}   \big)^{{d \over 2}+g_\sigma}}
The momentum integral must be performed in components so the sum is complicated; $D'_\sigma$ are some messy coefficients,
which can be functions of $y=\beta /(\gamma+\beta)$ and $k_i$; $g_\sigma$ are integer-valued. Looking at the gaussian
formula, one can be convinced, that the form of the exponentials remains exactly the
same for each term as in the scalar case. Therefore, as in the scalar case, we're justified in associating the
masses with the coefficients of the $y$ powers in the exponent of the exponentials.

\listrefs

\end